\documentclass[preprint,12pt]{elsarticle}
\usepackage{bm}
\usepackage{amsfonts}
\usepackage{graphicx}
\usepackage{mathrsfs}
\usepackage{amsmath}
\usepackage{amssymb}
\usepackage{float}
\newtheorem{thm}{Theorem}
\newtheorem{lemma}{Lemma}

\def\var{\mathrm {var}}

\def\bm{\boldsymbol}

\def\tr{\mathrm {tr}}

\def\U{{\bf U}}
\def\A{{\bf A}}
\def\M{{\bf M}}

\def\Z{{\bm Z}}

\def\S{{\bf S}}

\def\I{{\bf I}}

\def\u{{\bf u}}

\def\X{{\bm X}}

\def\Y{{\bm Y}}
\def\V{{\bm V}}

\def\tr{\mathrm {tr}}

\def\bmv{\bm \varepsilon}
\def\bth{{\bm\theta}}
\def\bms{{\bm\Sigma}}
\def\bmo{{\bm \omega}}

\def\cp{\mathop{\rightarrow}\limits^{p}}
\def\cd{\mathop{\rightarrow}\limits^{d}}

\usepackage{color}

 \biboptions{comma,round}

\journal{Statistics $\&$ Probability Letters}

\begin{document}

\begin{frontmatter}


\title{ Power Comparison between High Dimensional $t$-Test, Sign, and Signed Rank Tests}
\author{Long Feng\corref{cor1}}
\address{Key Laboratory for Applied Statistics of MOE and School of Mathematics and Statistics, Northeast Normal University, Changchun 130024, Jilin Province, China}
\cortext[cor1]{Corresponding author: flnankai@126.com}

\begin{abstract}
In this paper, we propose a power comparison between high
dimensional $t$-test, sign and signed rank test for the one sample
mean test. We show that the high dimensional signed rank test is
superior to a high dimensional $t$ test, but inferior to a high
dimensional sign test.
\end{abstract}



\end{frontmatter}

\section{Introduction}
Testing the population mean vector is a fundamental problem in
statistics.  For univariate data, the classic $t$ test is a very
popular method. However, it is not robust and is sensitive to
outliers and heavy tailed distributions, and the sign and Wilcoxon
signed rank tests are preferable for heavy tailed distributions. For
multivariate data, Hotelling's $T^2$ test is a natural extension of
the $t$ test, but is also not robust for heavy tailed distributions.
Randles (2000) extended the sign test to a multivariate sign test
for elliptical symmetric distributions, and
M$\ddot{o}$tt$\ddot{o}$nen and Oja (1995) proposed a multivariate
signed rank test and showed it to be very efficient. With the rapid
development of technology, various high dimensional datasets have
been generated in many areas, such as hyperspectral imagery,
internet portals, microarray analysis, and DNA mapping. This article
considers high dimensional testing of location parameters where the
dimensionality is potentially much larger than the sample size.

In multivariate hypothesis testing, a good estimator of the scatter
matrix  is a very important for the affine invariant test statistic.
However, the sample covariance matrix is not invertible when the
dimension, $p$, exceeds the sample size, $n$, and classic
multivariate tests cannot be applied. Recent research has focused on
large scale covariance matrix estimation under certain sparseness
conditions (Bickel and Levina, 2008; Cai and Liu, 2011). While it
seems reasonable to replace the scatter matrix with sparse matrix
estimators, it is difficult to maintain significance levels for
those modified test statistics (Feng, Zou, and Wang, 2016). An
alternative method is to replace the sample scatter matrix by its
diagonal matrix or the identity matrix, and many modified
Hotelling's $T^2$ tests have been proposed based on this approach,
such as Bai and Saranadasa (1996); Chen and Qin (2010); and Feng et
al. (2016). However, these methods are all based on the diverging
factor model or multivariate normal distributions, and are not
efficient for heavy tailed distributions. Wang et al. (2015)
proposed a high dimensional nonparametric multivariate test based on
multivariate sign for the one sample location problem. Feng, Zou and
Wang (2016) also proposed a scalar invariant high dimensional sign
test for the two sample location problem. They demonstrated that the
multivariate sign is a very efficient method to construct a robust
test for high dimensionality.

Multivariate signed rank (M$\ddot{o}$tt$\ddot{o}$nen and Oja, 1995)
is  another efficient method to construct robust multivariate
statistics, but also cannot be used in high dimensionality. We
propose a high dimensional signed rank test (SR), replacing the
sample scatter matrix in the multivariate signed rank test statistic
with the identity matrix. The proposed SR test statistic is shown to
be asymptotically normal under elliptical distributions, and
simulation comparisons show the SR procedure performs reasonably
well in terms of size and power for a wide range of dimensions,
sample sizes, and distributions.

For the main contribution of our paper, we derive the explicit form
of the asymptotic relative efficiency (ARE) for the SR test relative
to the high dimensional $t$ test (Chen and Qin, 2010) and high
dimensional sign test (Wang, Peng, and Li, 2015). For univariate or
multivariate data, the sign and SR tests perform better than the $t$
test for heavy tailed distributions. For high dimensional data, the
high dimensional sign test and proposed SR test are not worse than
the high dimensional $t$-test. For light tailed distributions, the
three tests are equivalent, but for heavy tailed distributions, the
high dimensional sign and proposed SR test will be more powerful
than high dimensional $t$ test, which we have verified by
simulation.

\section{High dimensional signed rank test}
Assume $\{\X_i\}_{i=1}^n$ are i.i.d. random samples from a $p$ variate elliptically symmetric distribution with density function
$\mbox{det}({\bf \Sigma})^{-1/2}g(||{\bf  \Sigma}^{-1/2}({\bf x}-\bm\theta)||)$,
where $\bm\theta$ are the symmetry centers, and ${\bf\Sigma}$ are the positive definite symmetric $p\times p$ scatter matrices.  Consider the one sample testing problem
\begin{align*}
H_0: \bth=\bm 0 ~~~\text{versus}~~~H_1: \bth\not=\bm 0.
\end{align*}
For univariate data, the Wilcoxon signed rank test statistic is essentially the sign test statistic applied to the Walsh sums (or averages), $x_i+x_j$ for $i\le j$. Similarly, the multivariate one sample signed rank test statistic can be constructed using the signs of transformed Walsh sums (or averages),
\begin{align*}
Q^2=\frac{np}{4c_1^2}\left|\left|n^{-2}\sum_{i=1}^n\sum_{j=1}^nU(\S^{-1/2}(\X_i+\X_j))\right|\right|^2,
\end{align*}
where $U({\bf x})=||{\bf x}||^{-1}{\bf x}I({\bf x}\neq {\bf 0})$, $\S^{-1/2}$ is the estimator for the scatter matrix, and $c_1$ is a scalar parameter. However, $\S$ is not available when the dimension is larger than the sample size, and $Q^2$ is not well defined. A natural extension is excluding $\S$ in $Q^2$,
\begin{align*}
\tilde{Q}^2=\frac{np}{4c_2^2}\left|\left|n^{-2}\sum_{i=1}^n\sum_{j=1}^nU(\X_i+\X_j)\right|\right|^2.
\end{align*}
However, the expectation of $\tilde{Q}^2$ is not zero and difficult to calculate. We also exclude the ``same" terms,  $U(\X_i+\X_j)^TU(\X_i+\X_k)$ and $U(\X_i+\X_j)^TU(\X_i+\X_j)$, and propose the SR test statistic
\begin{align*}
T_n=\frac{1}{P_n^4}\sum^{*} U(\X_i+\X_j)^TU(\X_k+\X_l),
\end{align*}
where $P_n^m={n!}/{(n-m)!}$; and $\sum\limits^{*}$ denotes summation over distinct indexes, e.g. the summation in $T_n$ is over the set $\{i\not=j\not=k\not=l\}$, for all $i,j,k,l\in\{1,\cdots,n\}$.

This appears to be an $O(n^4p)$ calculation for $T_n$, but $O(n^3p)$ is sufficient, since
\begin{align*}
T_n=\frac{1}{P_n^4}\left(W_1-W_2-2n(n-1)\right),
\end{align*}
where
\begin{align*}
W_1=\left|\left|\sum^{*}U(\X_i+\X_j)\right|\right|^2,~~~
W_2=\sum^*U(\X_i+\X_j)^TU(\X_i+\X_k).
\end{align*}
There are $O(n^2p)$ calculations for $W_1$ and $O(n^3p)$ for $W_2$. Thus, we need only $O(n^3p)$ calculations for $T_n$.

The following conditions are required for asymptotic analysis.
\begin{enumerate}
\item[(C1)] $\tr(\bms^4)=o(\tr^2(\bms^2))$.
\item[(C2)] $\tr(\bms^2)-p=o(n^{-1}p^2)$.
\end{enumerate}
(C1) is similar to condition (3.8) in Chen and Qin (2010), and will
hold if all the eigenvalues of $\bms$ are bounded. Define
$\bmv=\bms^{-1/2}(\X-\bth)$. Similar to Condition (C4) in Feng, Zou
and Wang (2016), (C2) is used to reduce the difference between
modules $||\bmv||$ and $||\bms^{1/2}\bmv||$. Thus, we may obtain an
explicit relationship between the variance of $T_n$ and $\bms$.
Consider a simple setting, $\tr(\bms^2)=O(p)$, (C2) becomes $p/n\to
\infty$.

\begin{thm}
Under conditions (C1), (C2), and $H_0$, as $(p,n)\to \infty$,
\begin{align*}
T_n/\sigma_n\cd N(0,1),
\end{align*}
where $\sigma_n^2=\frac{8\tr(\bms^2)}{n^2p^2}$.
\end{thm}

We propose a ratio consistent estimator for $\tr(\bms^2)$ (Feng, 2015),
\begin{align*}
\widehat{\tr(\bms^2)}=\frac{2p^2}{P_n^4}\sum^{*}
U(\X_i-\X_j)^TU(\X_k-\X_l)U(\X_k-\X_j)^TU(\X_i-\X_l),
\end{align*}
which requires $O(n^4p)$ calculations. However, one can simply fix  $i=n/2$, so that $\widehat{\tr(\bms^2)}$ requires only $O(n^3p)$ calculations. Simulation studies show this modified estimator is similar to the native estimator.

Consider the asymptotic distribution of $T_n$ under the alternative hypothesis
\begin{enumerate}
\item[(C3)] $\bth^T\bth=O(c_0^{-2}\sigma_n)$, $\bth^T\bms\bth=o(npc_0^{-2}\sigma_n)$,  where $c_0=E(||\X_{i}+\X_{j}||^{-1})$.
\end{enumerate}
(C3) requires that the difference between $\bth$ and $\bm 0$ is not
large such that the variance of $T_n$ is still asymptotic to
$\sigma_n^2$. This can be viewed as a high dimensional version of
the local alternative hypotheses.
\begin{thm}
Under conditions (C1)--(C3), as $(p,n)\to \infty$,
\[\frac{T_n-4c_0^2\bth^T\bth}{\sigma_n}\cd N(0,1).\]
\end{thm}
Consequently, the SR asymptotic power becomes
\begin{align*}
\beta_{\rm
SR}(||\bth||)&=\Phi\left(-z_{\alpha}+\frac{2c_0^2pn\bth^T\bth}{\sqrt{2\tr({\bms}^2)}}\right).
\end{align*}

\section{Power comparison}
Chen and Qin (2010) proposed a high dimensional two sample location (CQ) test by modifying the classic Hotelling's $T^2$ test. We simplify this to the one sample problem, i.e.,
\begin{align*}
T_{CQ}=\frac{1}{n(n-1)}\sum^{*}\X_i^T\X_j.
\end{align*}
They showed that the asymptotic power the CQ test was
\begin{align*}
\beta_{\rm
CQ}(||\bth||)&=\Phi\left(-z_{\alpha}+\frac{np\bth^T\bth}{E(||\bm
\varepsilon||^2)\sqrt{2\tr({\bms}^2)}}\right).
\end{align*}
The SR (ARE) with respect to CQ is
\begin{align*}
{\rm ARE}({\rm SR}, {\rm CQ})=2c_0^2E(||\bm\varepsilon||^2).
\end{align*}
Since
$E(||(\bms^{1/2}-\I_p)(\bmv_{i}+\bmv_{2j})||^2)=O(\tr(\bms^{1/2}-\I_p)^2)=o(n^{-1}p^2)$,
\begin{align*}
||\bms^{1/2}(\bmv_{i}+\bmv_{j})||=&||(\bmv_{i}+\bmv_{j})+(\bms^{1/2}-\I_p)(\bmv_{i}+\bmv_{j})||\\
=&||\bmv_{i}+\bmv_{j}||(1+o_p(1)),
\end{align*}
and $c_0=E(||\bm\varepsilon_{1}+\bmv_2||^{-1})(1+o(1))$. So,
\begin{align*}
{\rm ARE}({\rm SR}, {\rm
CQ})\approx& 2\{E(||\bm\varepsilon_{1}+\bmv_2||^{-1})\}^2E(||\bm \varepsilon||^2)\\
=&\{E(||\bm\varepsilon_{1}+\bmv_2||^{-1})\}^2E(||\bm
\varepsilon_1+\bm \varepsilon_2||^2)\\
\ge &\{E(||\bm\varepsilon_{1}+\bmv_2||^{-1})E(||\bm
\varepsilon_1+\bm \varepsilon_2||)\}^2 \ge 1,
\end{align*}
Thus, the proposed SR test can be no worse than the CQ test. From
the Cauchy inequality, ${\rm ARE}({\rm SR}, {\rm CQ})$ is one if and
only if $\frac{\var(||\bmv_1+\bmv_2||)}{E(||\bmv_1+\bmv_2||^2)}\to
0$, or $\frac{||\bmv_1+\bmv_2||^2}{E(||\bmv_1+\bmv_2||^2)}\cp 1$.
Therefore, if $||\bmv_1+\bmv_2||^2$ converges to
$E(||\bmv_1+\bmv_2||^2)$, the SR test is asymptotically equivalent
to CQ. For example, if $\bmv \sim N(\bm 0,\I_p)$,
$||\bmv_1+\bmv_2||/\sqrt{2p}\cp 1$, and SR has the same efficiency
as CQ. Otherwise, SR is preferable in terms of asymptotic power
under local alternatives, since  ${\rm ARE}({\rm SR}, {\rm CQ})$
increases with increasing $||\bmv_1+\bmv_2||$ variance.

Wang et al. (2015) proposed a high dimensional one sample location (SS) test based on spatial sign,
\begin{align*}
T_{SS}=\frac{1}{n(n-1)}\sum^{*}U(\X_i)^TU(\X_j).
\end{align*}
They showed the ARE of their SS test with respect to CQ was
\begin{align*}
{\rm ARE}({\rm SS}, {\rm
CQ})=&\{E(||\bm\varepsilon||^{-1})\}^2E(||\bm \varepsilon||^2)\\
\ge &\{E(||\bm\varepsilon||^{-1})E(||\bm \varepsilon||)\}^2 \ge 1.
\end{align*}
Similarly,  if $||\bmv||$ converges to $E(||\bmv||)$, SS is
asymptotically equivalent to CQ. Otherwise,  SS will be more
powerful than CQ, since ${\rm ARE}({\rm SS}, {\rm CQ})$ increases
with increasing $||\bmv||$ variance.

Therefore, the SR ARE with respect to SS is
\begin{align*}
{\rm ARE}({\rm SR}, {\rm
SS})=\frac{2\{E(||\bm\varepsilon_{1}+\bmv_2||^{-1})\}^2}{\{E(||\bm\varepsilon||^{-1})\}^2}
\end{align*}
Define $\bmv_i=R_i\U_i$, where $R_i=||\bmv_i||$, $\U_i=U(\bmv_i)$,
then
$||\bmv_1+\bmv_2||^2=R_1^2+R_2^2+2R_1R_2\U_1^T\U_2$,$E((R_1R_2\U_1^T\U_2)^2)=p^{-2}E(R_1^2)E(R_2^2)$.
So $||\bmv_1+\bmv_2||^2=(R_1^2+R_2^2)(1+o_p(1))$ and
$||\bmv_1+\bmv_2||^{-1}=(||\bmv_1||^2+||\bmv_2||^2)^{-1/2}(1+o_p(1))$.
\begin{align*}
{\rm ARE}({\rm SR}, {\rm SS})&\approx\frac{2\{E((||\bmv_1||^2+||\bmv_2||^2)^{-1/2})\}^2}{\{E(||\bm\varepsilon||^{-1})\}^2}\\
&\le
\frac{2\{E(||\bmv_1||^{-1}+||\bmv_2||^{-1})/\sqrt{2}\}^2}{\{E(||\bm\varepsilon||^{-1})\}^2}=1.
\end{align*}
The relation holds only with $||\bmv||^2/E(||\bmv||^2)\cp 1$. Otherwise, SS is more efficient than SR, since ${\rm ARE}({\rm SR}, {\rm SS})$ increases with increasing $||\bmv||$ variance.

Generally, if $||\bmv||^2/E(||\bmv||^2)\cp 1$, these three tests are equivalent. Otherwise, SS is superior, then SR and finally CQ. Table \ref{t1} shows the ARE between the tests under multivariate $t$ distributions with different degrees of freedom and mixed normal distributions.

 \begin{table}[ht]
           \centering
           \caption{Asymptotic relative efficiencies for different distributions. ($t_{\nu}$ denotes multivariate $t$ distributions with degree of freedom $\nu$; $MN(\gamma,\tau)$ denotes mixed normal distributions with density function $\gamma N(\bm 0,\I_p)+(1-\gamma)N(\bm 0,\tau^2\I_p)$.)}
           \vspace{0.1cm}
      \renewcommand{\arraystretch}{1.1}
     \tabcolsep 7pt
         \begin{tabular}{ccccccccc}\hline \hline
  & $t_3$ & $t_4$ &$t_5$ &$t_6$ & $t_{10}$ & $N(\bm 0, \I_p)$ &MN(0.2,3) &MN(0.05,10)\\
  {\rm ARE(SS,CQ)} & 2.54 & 1.76 & 1.51 & 1.38 &1.18 & 1.00 &1.95 &5.79\\
  {\rm ARE(SR,CQ)} & 1.98 & 1.48 & 1.31 & 1.22 &1.10 & 1.00 &1.64 &5.26\\
  {\rm ARE(SR,SS)} & 0.78 & 0.84 & 0.87 & 0.88 &0.93 & 1.00 &0.84 &0.91\\ \hline \hline
               \end{tabular}\label{t1}\\
               \vspace{0.1cm}
  {\small Note: These results were calculated by simulation with $p=2000$ and 10,000 replications.}
           \end{table}

\section{Simulation}
A simulation study was undertaken to evaluate the performance of the proposed SR test, incorporating 2,500 replications. Three scenarios were considered.
\begin{enumerate}
\item[(I)] Multivariate normal distribution. $\X_i\sim N(\bm\theta,\bms)$.
\item[(II)] Multivariate $t$ distribution $t_{p,4}$.   $\X_{i}$'s were generated from $t_{p,4}$ with $\bms$.
\item[(III)] Multivariate mixed normal distribution $\mbox{MN}_{p,\gamma,9}$. $\X_{i}$'s were generated from $(1-\gamma) f_p(\bm\theta,\bms)+\gamma f_p(\bm\theta,9\bms)$, denoted by $\mbox{MN}_{p,\gamma,9}$, where $f_p(\cdot;\cdot)$ is the density function of $p$-variate multivariate normal distribution. $\gamma$ was chosen to be 0.2.
\end{enumerate}
The scatter matrix was $\bms=(0.5^{|i-j|})$. First, consider the low dimensional case $p<n$. Two sample sizes $n=30,40$ were evaluated with dimension $p=0.8n$. Under the alternative hypothesis, two allocation patterns were considered: dense and sparse. These assumed the first $50\%$ and $95\%$, respectively, of components of $\bth$ to be zero. We set $\bth^T\bth/\sqrt{\tr({\bf \Lambda}^2)}=0.1$, where ${\bf \Lambda}$ is the covariance matrix. Since the empirical sizes of the classic spatial signed rank test (TSR) deviate significantly from the nominal, we propose a size corrected power comparison. The critical values of TSR and SR tests were found through simulation, so both tests have accurate sizes for each case. Table \ref{t2} shows the size corrected power of the tests. The proposed SR test is more powerful than TSR when dimension is comparable to sample size. However, classical Mahalanobis distance may not work well because the contamination bias in estimating the covariance matrix grows rapidly with $p$ (Bai and Saranadasa, 1996). When $p$ and $n$ are comparable, using the inverse of the estimate of the scatter matrix in constructing tests is no longer beneficial.

 \begin{table}[ht]
           \centering
           \caption{Size corrected  power comparison at 5\% significance for low dimensionality ($p<n$).}
           \vspace{0.1cm}
      \renewcommand{\arraystretch}{1.2}
     \tabcolsep 7pt
         \begin{tabular}{ccccccccccccccc}\hline \hline
     &  \multicolumn{4}{c}{Dense Case}&  &  \multicolumn{4}{c}{Sparse Case}     \\
$(n_i,p)$  & \multicolumn{2}{c}{(30,24)} & \multicolumn{2}{c}{(40,32)} & & \multicolumn{2}{c}{(30,24)} & \multicolumn{2}{c}{(40,32)} \\
Scenario     & TSR &SR & TSR &SR & & TSR &SR & TSR &SR  \\
(I)  & 9.6 &51.8  &11.4  &65.4  & &11.8 &86.0 &16.2 &86.8 \\
(II) &11.7 &68.2  &16.6  &84.8  & &16.0 &97.0 &21.5 &96.8\\
(III)&15.2 &74.0  &19.6  &88.4  & &17.2 &97.8 &25.5 &97.9\\
 \hline \hline
               \end{tabular}\label{t2}
           \end{table}

Consider the high dimensional case, $p>n$. Two sample sizes $n=30,40$ and three dimensions $p=100,200,400$ were evaluated with $\bth^T\bth/\sqrt{\tr({\bf \Lambda}^2)}=0.05$. The other settings were the same as the low dimensional case. Table \ref{t3} shows the comparison between SR, CQ, and SS tests for empirical size and power. All tests maintain significance well in all cases. For scenario I, SS and SR tests perform similarly to CQ. For multivariate normal distributions, the difference between direction and moment based tests disappears with increasing dimensionality. However, when the underlying distribution is not normal, direction based tests are superior to moment based tests. For scenarios II and III, both SS and SR tests are significantly superior to CQ, and asymptotic relative efficiencies are close to the theoretical results from Table \ref{t1}. As expected, the proposed SR test outperforms CQ, but is slightly inferior to SS for multivariate $t$ and mixed normal distributions.

 \begin{table}
                     \centering
                     \caption{Empirical sizes and power ($\%$)at 5\% significance for scenarios I--III}
                     \vspace{0.1cm}
                \renewcommand{\arraystretch}{1.0}
               \tabcolsep 7pt{
                   \begin{tabular}{cccccccccccccc}\hline \hline
                   && \multicolumn{3}{c}{Size} && \multicolumn{3}{c}{Dense} &&\multicolumn{3}{c}{Sparse}\\ \cline{3-5} \cline{7-9}\cline{11-13}
 $n$ & $p$ &CQ &SS & SR &&CQ &SS & SR &&CQ &SS & SR\\
 \multicolumn{13}{c}{Scenario I} \\ \hline
30 &100  &  4.7 &  5.6 &  5.7  && 26.9 & 30.0 & 29.6  && 33.2 & 37.4 & 36.4 \\
30 &200  &  5.0 &  6.8 &  6.3  && 26.4 & 29.9 & 29.3  && 29.8 & 31.9 & 32.8 \\
30 &400  &  4.8 &  6.0 &  5.8  && 26.3 & 30.0 & 29.3  && 29.0 & 32.5 & 31.9 \\
40 &100  &  4.5 &  5.8 &  4.8  && 38.0 & 40.2 & 39.6  && 46.4 & 49.1 & 48.0 \\
40 &200  &  4.9 &  6.2 &  5.8  && 38.5 & 41.2 & 41.1  && 42.6 & 45.1 & 45.4 \\
40 &400  &  5.3 &  6.2 &  6.2  && 37.1 & 40.5 & 40.4  && 41.9 & 44.4 & 43.7 \\
   \multicolumn{13}{c}{Scenario II} \\ \hline
30 &100  &  5.5 &  5.6 &  5.2  && 31.2 & 50.5 & 42.7  && 39.6 & 63.0 & 55.0 \\
30 &200  &  4.5 &  6.8 &  5.9  && 31.1 & 55.4 & 44.9  && 34.8 & 61.4 & 50.2 \\
30 &400  &  4.5 &  6.0 &  5.6  && 29.0 & 56.2 & 43.8  && 32.1 & 59.5 & 48.9 \\
40 &100  &  4.3 &  5.8 &  5.2  && 41.3 & 67.7 & 58.4  && 49.1 & 80.1 & 67.8 \\
40 &200  &  5.0 &  6.2 &  5.5  && 42.7 & 69.6 & 59.6  && 48.8 & 77.3 & 67.0 \\
40 &400  &  5.8 &  6.2 &  6.5  && 41.7 & 72.2 & 62.7  && 45.7 & 75.7 & 65.5 \\
 \multicolumn{13}{c}{Scenario III} \\ \hline
30 &100  &  5.1 &  5.6 &  5.6  && 29.7 & 56.0 & 47.5  && 36.8 & 68.6 & 59.8 \\
30 &200  &  4.9 &  6.8 &  5.8  && 29.3 & 60.3 & 48.8  && 32.8 & 67.9 & 56.6 \\
30 &400  &  4.8 &  6.0 &  5.3  && 29.6 & 62.3 & 53.8  && 31.2 & 64.5 & 55.4 \\
40 &100  &  5.0 &  5.8 &  5.0  && 39.4 & 72.2 & 62.9  && 45.7 & 85.1 & 74.9 \\
40 &200  &  4.7 &  6.2 &  6.4  && 42.8 & 75.8 & 66.1  && 45.7 & 82.2 & 71.9 \\
40 &400  &  4.0 &  6.2 &  5.5  && 39.8 & 77.9 & 68.8  && 44.3 &
81.4& 72.2\\ \hline \hline
                    \end{tabular}}\label{t3}
                \end{table}

To give a broader picture of the robustness and efficiency of the proposed method, we also considered the diverging factor models (Bai and Saranadasa, 1996), $\X_i=\bm\theta+\bms^{1/2}\Z_i$ where $\Z_i=(Z_{i1},\cdots,Z_{ip})$. Two $Z_{ip}$ distributions were considered,
\begin{enumerate}
\item[(IV)] $Z_{ip}\sim t_4$.
\item[(V)]  $Z_{ip}\sim 0.8N(0,1)+0.2N(0,9)$.
\end{enumerate}
Random vectors from scenarios IV and V are not elliptically distributed, and Table \ref{t4} shows the simulated results for these scenarios with the same settings as Table \ref{t3}. SR controls empirical size very well in these cases. SR test power is also lager than CQ and a little smaller than SS. Thus, the proposed SR test performs very well for non-elliptical distributions.

The outcomes verify the proposed SR test is efficient and robust across a wide range of distributions. When the distributions are heavy tailed, SR is significantly more efficient than moment based tests. This may be due to the proposed test using only the observation direction from the origin, rather than its distance from the origin, which would tend to be more for heavy tailed distributions.

 \begin{table}[ht]
                     \centering
                     \caption{Empirical sizes and power ($\%$)at 5\% significance for scenarios IV--V}
                     \vspace{0.1cm}
                \renewcommand{\arraystretch}{1.0}
               \tabcolsep 7pt{
                   \begin{tabular}{cccccccccccccc}\hline \hline
                   && \multicolumn{3}{c}{Size} && \multicolumn{3}{c}{Dense} &&\multicolumn{3}{c}{Sparse}\\ \cline{3-5} \cline{7-9}\cline{11-13}
 $n$ & $p$ &CQ &SS & SR &&CQ &SS & SR &&CQ &SS & SR\\
 \multicolumn{13}{c}{Scenario IV} \\ \hline
30 &100  &  5.6 &  7.4 &  6.1 & & 26.6 & 30.4 & 29.1&  & 31.5 & 37.2 & 36.4\\
30 &200  &  3.6 &  5.8 &  5.7 & & 27.5 & 31.0 & 30.5&  & 30.8 & 33.4 & 33.3\\
30 &400  &  4.5 &  5.5 &  5.7 & & 24.9 & 30.0 & 29.2&  & 26.7 & 31.3 & 29.8\\
40 &100  &  5.7 &  7.3 &  6.3 & & 39.5 & 42.7 & 41.6&  & 46.9 & 51.9 & 50.1\\
40 &200  &  5.6 &  6.9 &  6.7 & & 35.5 & 38.6 & 38.0&  & 38.3 & 42.1 & 41.5\\
40 &400  &  6.0 &  6.9 &  6.9 & & 39.3 & 43.1 & 42.8&  & 41.2 & 44.1 & 44.1\\
   \multicolumn{13}{c}{Scenario V} \\ \hline
30 &100  &  4.5 &  7.1 &  6.0 & & 29.5 & 55.4 & 46.6&  & 33.9 & 66.1 & 56.7\\
30 &200  &  5.7 &  6.1 &  6.0 & & 30.1 & 58.6 & 48.5&  & 34.3 & 63.5 & 53.3\\
30 &400  &  3.7 &  6.4 &  5.3 & & 30.9 & 58.8 & 49.8&  & 30.8 & 62.6 & 51.7\\
40 &100  &  6.5 &  7.1 &  6.0 & & 42.2 & 73.4 & 65.4&  & 50.4 & 84.0 & 74.5\\
40 &200  &  6.3 &  7.0 &  6.2 & & 43.2 & 76.4 & 66.3&  & 45.4 & 80.8 & 70.7\\
40 &400  &  4.6 &  4.9 &  4.5 & & 39.0 & 75.6 & 65.7&  & 42.4 & 79.4 & 69.7\\
\hline \hline
                    \end{tabular}}\label{t4}
                \end{table}

\section{Discussion}
Many high dimensional scalar invariant tests have been proposed (Park and Ayyala, 2013; Srivastava, 2009; Feng, et al., 2015; Feng, 2015). The concept is to replace $\S$ by its diagonal matrix in $Q^2$, so that all variables have the same scale. How to construct an SR scalar invariant test will be the topic of further study.

\section*{Acknowledgement}

This research was supported by the National Natural Science Foundation of China (grant No. 11501092) and the Fundamental Research Funds for the Central Universities.

\section{Appendix: Theorem Proofs}

Define $\u_{i}=E(U(\bmv_{i}-\bmv_{j})|\bmv_{i})$ and $E(\u_{i}\u_{i}^T)=\tau_Fp^{-1}\I_p$, where the constant $\tau_F$ depends on the background distribution, $F$;$\Y_{i}=\X_{i}-\bth$, $\V_{i}=E(U(\Y_{i}+\Y_{j})|\Y_{i})$; and $\bmo_{ij}=U(\Y_{i}+\Y_{j})-\V_{i}-\V_{j}$. Recall Zou, et al., 2014, Lemma 4:
\begin{lemma} \label{le1}
Suppose $\u$ are i.i.d. uniform on the unit $p$ sphere. For any $p\times p$ symmetric matrix $\M$,
\begin{align*}
E(\u^T\M\u)^2=&\{\tr^2(\M)+2\tr(\M^2)\}/(p^2+2p),\\
E(\u^T\M\u)^4=&\{3\tr^2(\M^2)+6\tr(\M^4)\}/\{p(p+2)(p+4)(p+6)\}.
\end{align*}
\end{lemma}

Recall, also, Feng, 2015, Lemma 3.
\begin{lemma}\label{le2}
$\tau_F\to 0.5$ as $p\to \infty$.
\end{lemma}

\subsection{Proof of Theorem 1}
From the above definitions and lemmas,
\begin{align*}
T_n=&\frac{1}{P_n^4}\sum^{*} U(\X_i+\X_j)^TU(\X_k+\X_l)\\
=&\frac{1}{P_n^4}\sum^{*}(\V_i+\V_j+P_{ij})^T(\V_k+\V_l+P_{kl})\\
=&\frac{4}{n(n-1)}\underset{i\not=j}{\sum\sum}\V_i^T\V_j+\frac{2}{P_n^3}\sum^{*}\bmo_{ij}^T\V_k+\frac{1}{P_n^4}\sum^{*}\bmo_{ij}^T\bmo_{kl}\\
\doteq &Z_n+J_1+J_2.
\end{align*}
We need to show that $J_1=o_p(\sigma_n), J_2=o_p(\sigma_n)$.

Under $H_0$, $E(\V_i)=0, E(\bmo_{ij})=0$, and from the definition of $\bmo_{ij}$, $E(V_i\bmo_{ij})=0$, $E(\bmo_{ij}\bmo_{ik})=0$. Therefore,
\begin{align*}
E(J_1^2)=O(n^{-3})E((\bmo_{ij}^T\V_k)^2)+O(n^{-3})E(\bmo_{ij}^T\V_i\bmo_{kj}^T\V_k).
\end{align*}
We need to show that $E((\bmo_{ij}^T\V_k)^2)=O(p^{-2}\tr(\bms^2))$ and $E(\bmo_{ij}^T\V_i\bmo_{kj}^T\V_k)=O(p^{-2}\tr(\bms^2))$.

Define $\A=E(\V_i\V_i^T)$, then $E((\bmo_{ij}^T\V_k)^2)=E(\bmo_{ij}^T\A \bmo_{ij})=E(U(\Y_i+\Y_j)^T\A U(\Y_i+\Y_j))-\tr(\A^2)$. From symmetry of $\Y_i$, and Lemma 1, $E(U(\Y_i+\Y_j)^T\A U(\Y_i+\Y_j))=E(U(\Y_i-\Y_j)^T\A U(\Y_i-\Y_j)) = O(\tr(\A^2))$. Similarly, $E(\bmo_{ij}^T\V_i\bmo_{kj}^T\V_k)=O(\tr(\A^2))$. Therefore,
\begin{align*}
\V_{i}=&E(U(\Y_{i}+\Y_{j})|\Y_{i})=E(U(\bms^{1/2}(\bmv_{i}+\bmv_{j}))|\bms^{1/2}\bmv_{i})\\
=&E(U(\bms^{1/2}(\bmv_{i}+\bmv_{j}))|\bmv_{i})\\
=&E\left(\frac{\bms^{1/2}(\bmv_{i}+\bmv_{j})}{||\bms^{1/2}(\bmv_{i}+\bmv_{j})||}\bigg|\bmv_{i}\right).
\end{align*}
Since $E(||(\bms^{1/2}-\I_p)(\bmv_{i}+\bmv_{2j})||^2)=O(\tr(\bms^{1/2}-\I_p)^2)=o(n^{-1}p^2)$,
\begin{align*}
||\bms^{1/2}(\bmv_{i}+\bmv_{j})||=&||(\bmv_{i}+\bmv_{j})+(\bms^{1/2}-\I_p)(\bmv_{i}+\bmv_{j})||\\
=&||\bmv_{i}+\bmv_{j}||(1+o_p(1)),
\end{align*}
and $\V_{i}=\bms^{1/2}\u_{i}(1+o_p(1))$. Therefore,
\begin{align*}
\tr(\A^2)=E((\V_i^T\V_j)^2)=E((\u_i\bms\u_j)^2)(1+o(1))=\tau_F^2p^{-2}\tr(\bms^2)(1+o(1)).
\end{align*}
From Lemma 2, $\tau_F\to 0.5$ as $p\to\infty$, so that$\tr(\A^2)=4^{-1}p^{-2}\tr(\bms^2)(1+o(1))$. Thus, $J_1=o_p(\sigma_n)$.

Similarly, $E(J_2)=o_p(\sigma_n)$.

Finally, we need to show that
\begin{align*}
Z_n/\sigma_n \cd N(0,1).
\end{align*}
Define $W_{nk}=\sum_{i=2}^kZ_{ni}$, where $Z_{ni}=\sum_{j=1}^{i-1}\frac{8}{n(n-1)}\V_i^T\V_j$, and let $\mathcal{F}_{n,i}=\sigma\{\V_1,\cdots,\V_i\}$ be the $\sigma$ field generated by $\{\V_j, j\le i\}$. Since $E(Z_{ni}|\mathcal{F}_{n,i-1})=0$, it follows that $\{W_{nk}, \mathcal{F}_{n,k}; 2\le k \le n\}$ is a zero mean martingale. The central limit theorem will hold if
\begin{align}\label{clt1}
\frac{\sum_{j=2}^{n}E[Z_{nj}^2|\mathcal{F}_{n,j-1}]}{\sigma_n^2}\cp
1,
\end{align}
and, for any $\epsilon>0$,
\begin{align}\label{clt2}
\sigma_n^{-2}\sum_{j=2}^{n}E[Z_{nj}^2I(|Z_{nj}|>
\epsilon\sigma_n|)|\mathcal{F}_{n,j-1}]\cp 0.
\end{align}
It can be shown that
\begin{align*}
\sum_{j=2}^{n}E(Z_{nj}^2|\mathcal{F}_{n,j-1})=&\frac{64}{n^2(n-1)^2}\sum_{j=2}^{n}\sum_{i=1}^{j-1}\V_i^T\A \V_i\\
&+\frac{64}{n^2(n-1)^2}\sum_{j=2}^{n}\underset{i_1<i_2}{\sum^{j-1}\sum^{j-1}}\V_{i_1}^T\A\V_{i_2}\\
\doteq& C_{n1}+C_{n2},
\end{align*}
and since $E(C_{n1})=\frac{32}{n(n-1)}\tr(\A^2)=\frac{8}{n(n-1)p^2}\tr(\bms^2)(1+o(1))$, then $\var(C_{n1})=O(n^{-5})\var((\V_i^T\A\V_i)^2)$. From Lemma 1, $\var((\V_i^T\A\V_i)^2)=O(\tr^2(\A^2)+\tr(\A^4))$. Thus, from C1, $\var(C_{n1})=O(n^{-5})\tr^2(\A^2)=o(\sigma_n^4)$. Thus, $C_{n1}/\sigma_n^2 \cp 1$.

Similarly, $E(C_{n2}^2)=O(n^{-4})\tr(\A^4)=o(\sigma_n^4)$.
Thus,  (\ref{clt1}) holds.

To prove (\ref{clt2}), by Chebyshev's inequality, we need only show that
\begin{align*}
E\left\{\sum_{j=2}^{n}E[Z_{nj}^4|\mathcal{F}_{n,j-1}]\right\}=o(\sigma_n^4).
\end{align*}
Note that
\begin{align*}
E\left\{\sum_{j=2}^{n}E[Z_{nj}^4|\mathcal{F}_{n,j-1}]\right\}=\sum_{j=2}^{n}E(Z_{nj}^4)
=O(n^{-8})\sum_{j=2}^{n}E\left(\sum_{i=1}^{j-1}\V_j^T\V_i\right)^4,
\end{align*}
which can be decomposed as $3Q+P$, where
\begin{align*}
Q=&O(n^{-8})\sum_{j=2}^n
\underset{s<t}{\sum^{j-1}\sum^{j-1}}E(\V_j^T\V_s\V_s^T\V_j\V_j^T\V_t\V_t^T\V_j)\\
P=&O(n^{-8})\sum_{j=2}^n\sum_{i=1}^{j-1}E((\V_j^T\V_i)^4).
\end{align*}
Since, $Q=O(n^{-5})E((\V_j^T\A\V_j)^2)=O(n^{-5})\tr^2(\A^2)$ from Lemma 1 and (C1), then $Q=o(\sigma_n^4)$. Similarly, $P=O(n^{-6})\tr^2(\A^2)=o(\sigma_n^4)$. This completes the proof.

\subsection{Proof of Theorem 2}

From Taylor's expansion,
\begin{align*}
U(\X_i+\X_j)=&U(2\bth+\Y_i+\Y_j)\\
=&U(\Y_i+\Y_j)+\frac{2}{||\X_i+\X_j||}(\I_p-U(\X_i+\X_j)U(\X_i+\X_j)^T)\bth+o_p(c_0^2\bth^T\bth)\\
=&U(\Y_i+\Y_j)+\frac{2}{||\X_i+\X_j||}\bth+o_p(\sigma_n).
\end{align*}
Then,
\begin{align*}
T_n=&\frac{1}{P_n^4}\sum^{*}U(\Y_i+\Y_j)^TU(\Y_k+\Y_l)+\frac{1}{P_n^4}\sum^{*}U(\Y_i+\Y_j)\frac{4}{||\X_k+\X_l||}\bth\\
&+\frac{1}{P_n^4}\sum^{*}\frac{4}{||\X_i+\X_j||||\X_k+\X_l||}\bth^T\bth+o_p(\sigma_n).
\end{align*}
Using the same procedure as Theorem 1,
\begin{align*}
T_n=\frac{4}{n(n-1)}\underset{i\not=j}{\sum\sum}\V_i^T\V_j+\frac{8}{n}\sum_{i=1}^nc_0\V_i^T\bth+4c_0^2\bth^T\bth+o_p(\sigma_n).
\end{align*}
Since
\begin{align*}
E\left(\frac{8}{n}\sum_{i=1}^nc_0\V_i^T\bth\right)^2=n^{-1}c_0^2\bth^T\A\bth=o(\sigma_n^2),
\end{align*}
then
$T_n=\frac{4}{n(n-1)}\underset{i\not=j}{\sum\sum}\V_i^T\V_j+4c_0^2\bth^T\bth+o_p(\sigma_n)$,
which follows from Theorem 1. This completes the proof.


\vspace{0.5cm} \noindent{\small\bf References} \footnotesize
\baselineskip 7pt
\begin{description}
\item Bai, Z.D., Saranadasa, H. 1996. Effect of high dimension: by an example of a two sample
problem. Statist. Sinica 6, 311--329.

\item Bickel, P.J.,  Levina, E. 2008. Regularized estimation of large covariance matrices. Ann. Statist. 36, 199--227.

\item Cai, T.,  Liu, W. 2011. Adaptive thresholding for sparse
covariance matrix estimation. J. Amer. Statist. Assoc. 106, 672--684.

\item Chen, S.X.,  Qin, Y.L. 2010. A two-sample test for
high-dimensional data with applications to gene-set testing. Ann. Statist. 38, 808--835.

\item Feng, L. 2015. High dimensional spatial rank test for two sample
location problem. arXiv: 1506.08315.

\item Feng, L., Zou, C., Wang, Z. 2016. Multivariate-sign-based
high-dimensional tests for the two-sample location problem. J. Amer.
Statist. Assoc. 111, 721--735.

\item Feng, L., Zou, C., Wang, Z., Zhu, L. 2016. Two-sample
Behrens-Fisher problem for high-dimensional data. Statist. Sinica
25, 1297--1312.

\item M$\ddot{o}$tt$\ddot{o}$nen, J., Oja, H. 1995. Multivariate spatial sign and
rank methods. J. Nonparametr. Stat. 5, 201--213.

\item Park, J., Ayyala, D.N. 2013. A test for the mean vector in
large dimension and small samples. J. Statist. Plann. Inference 143, 929--943.

\item Randles, R.H. 2000. A simpler, affine-invariant, multivariate,
distribution-free sign test. J. Amer. Statist. Assoc. 95, 1263--1268.

\item Srivastava, M.S. 2009. A test for the mean vector with fewer
observations than the dimension under non-normality. J. Multivariate Anal. 100, 518--532.

\item Wang, L., Peng, B., Li, R. 2015. A high-dimensional
nonparametric multivariate test for mean vector. J. Amer. Statist. Assoc. 110, 1658--1669.

\item Zou, C., Peng, L., Feng, L., Wang, Z. 2014. Multivariate
sign-based high-dimensional tests for sphericity. Biometrika 101,
229--236.

\end{description}

\end{document}